\documentclass[conference]{IEEEtran}

\IEEEoverridecommandlockouts
\usepackage{cite}
\usepackage{amsmath,amssymb,amsfonts}
\usepackage{algorithmic}
\usepackage{graphicx}
\usepackage{textcomp}
\usepackage{xcolor}

\DeclareMathAlphabet{\mathsfbr}{OT1}{cmss}{m}{n}
\SetMathAlphabet{\mathsfbr}{bold}{OT1}{cmss}{bx}{n}
\DeclareRobustCommand{\msf}[1]{%
  \ifcat\noexpand#1\relax\msfgreek{#1}\else\mathsfbr{#1}\fi
}

\makeatletter
\newcommand{\msfgreek}[1]{\csname s\expandafter\@gobble\string#1\endcsname}
\makeatother

\DeclareFontEncoding{LGR}{}{} 
\DeclareSymbolFont{sfgreek}{LGR}{cmss}{m}{n}
\SetSymbolFont{sfgreek}{bold}{LGR}{cmss}{bx}{n}
\DeclareMathSymbol{\salpha}{\mathord}{sfgreek}{`a}
\DeclareMathSymbol{\sbeta}{\mathord}{sfgreek}{`b}
\DeclareMathSymbol{\sgamma}{\mathord}{sfgreek}{`g}
\DeclareMathSymbol{\sdelta}{\mathord}{sfgreek}{`d}
\DeclareMathSymbol{\sepsilon}{\mathord}{sfgreek}{`e}
\DeclareMathSymbol{\szeta}{\mathord}{sfgreek}{`z}
\DeclareMathSymbol{\seta}{\mathord}{sfgreek}{`h}
\DeclareMathSymbol{\stheta}{\mathord}{sfgreek}{`j}
\DeclareMathSymbol{\siota}{\mathord}{sfgreek}{`i}
\DeclareMathSymbol{\skappa}{\mathord}{sfgreek}{`k}
\DeclareMathSymbol{\slambda}{\mathord}{sfgreek}{`l}
\DeclareMathSymbol{\smu}{\mathord}{sfgreek}{`m}
\DeclareMathSymbol{\snu}{\mathord}{sfgreek}{`n}
\DeclareMathSymbol{\sxi}{\mathord}{sfgreek}{`x}
\DeclareMathSymbol{\somicron}{\mathord}{sfgreek}{`o}
\DeclareMathSymbol{\spi}{\mathord}{sfgreek}{`p}
\DeclareMathSymbol{\srho}{\mathord}{sfgreek}{`r}
\DeclareMathSymbol{\ssigma}{\mathord}{sfgreek}{`s}
\DeclareMathSymbol{\stau}{\mathord}{sfgreek}{`t}
\DeclareMathSymbol{\supsilon}{\mathord}{sfgreek}{`u}
\DeclareMathSymbol{\sphi}{\mathord}{sfgreek}{`f}
\DeclareMathSymbol{\schi}{\mathord}{sfgreek}{`q}
\DeclareMathSymbol{\spsi}{\mathord}{sfgreek}{`y}
\DeclareMathSymbol{\somega}{\mathord}{sfgreek}{`w}

\DeclareMathSymbol{\svarsigma}{\mathord}{sfgreek}{`c}

\DeclareMathSymbol{\sGamma}{\mathalpha}{sfgreek}{`G}
\DeclareMathSymbol{\sDelta}{\mathalpha}{sfgreek}{`D}
\DeclareMathSymbol{\sTheta}{\mathalpha}{sfgreek}{`J}
\DeclareMathSymbol{\sLambda}{\mathalpha}{sfgreek}{`L}
\DeclareMathSymbol{\sXi}{\mathalpha}{sfgreek}{`X}
\DeclareMathSymbol{\sPi}{\mathalpha}{sfgreek}{`P}
\DeclareMathSymbol{\sSigma}{\mathalpha}{sfgreek}{`S}
\DeclareMathSymbol{\sUpsilon}{\mathalpha}{sfgreek}{`U}
\DeclareMathSymbol{\sPhi}{\mathalpha}{sfgreek}{`F}
\DeclareMathSymbol{\sPsi}{\mathalpha}{sfgreek}{`Y}
\DeclareMathSymbol{\sOmega}{\mathalpha}{sfgreek}{`W}

\DeclareRobustCommand{\mcal}[1]{%
  \ifcat\noexpand#1\relax\mathnormal{#1}\else\cal{#1}\fi
}
\DeclareRobustCommand{\BM}[1]{%
  \ifcat\noexpand#1\relax\bm{\boldUppercaseItalicGreek{#1}}\else\bm{#1}\fi
}
\makeatletter
\newcommand{\boldUppercaseItalicGreek}[1]{\csname var\expandafter\@gobble\string#1\endcsname}
\makeatother

\newcommand{\V}[1]{\bm{#1}} 
\newcommand{\Set}[1]{{\mcal{#1}}} 

\newcommand{\E}[1]{\mathbb{E}\left\{#1\right\}}

\newcommand{\avg}[1]{\overline{\left\{#1\right\}}}

\usepackage{bm}
\usepackage{color, courier}
\usepackage{psfrag}
\usepackage{acronym}
\usepackage{amsmath}
\usepackage{amssymb}
\usepackage{amsfonts}
\usepackage[colorlinks = true, allcolors = black]{hyperref}
\usepackage{latexsym}
\usepackage{mathrsfs}
\usepackage{epstopdf}
\usepackage{algorithm}
\usepackage{mathtools}
\usepackage[caption=false, font=scriptsize]{subfig}
\usepackage{algorithmic}
\usepackage{relsize}

\DeclareMathOperator*{\argmax}{arg\,max}

\newcommand{\st}{\operatorname{s.t.}\,}

\newtheorem{definition}{Definition}

\newtheorem{proposition}{Proposition}

\newtheorem{corollary}{Corollary}
\newtheorem{remark}{Remark}

\definecolor{green}{rgb}{0, 0.5, 0}
\definecolor{pink}{rgb}{1, 0, 1}

\acrodef{agi}[AgI]{augmented information}
\acrodef{mec}[MEC]{mobile edge computing}
\acrodef{ldp}[LDP]{Lyapunov drift-plus-penalty}
\acrodef{lp}[LP]{linear programming}
\acrodef{cmdp}[CMDP]{constrained Markov decision process}
\acrodef{pdf}[pdf]{probability density function}
\acrodef{ue}[UE]{user equipment}
\acrodef{sfc}[SFC]{service function chain}
\acrodef{bp}[BP]{back-pressure}

\graphicspath{{figures/}}

\begin{document}

\title{
Optimal Cloud Network Control with \\
Strict Latency Constraints
}

\author{
	\IEEEauthorblockN{Yang Cai\IEEEauthorrefmark{1}, Jaime Llorca\IEEEauthorrefmark{2}, Antonia M. Tulino\IEEEauthorrefmark{2}\IEEEauthorrefmark{3}, Andreas F. Molisch\IEEEauthorrefmark{1}} 
	\IEEEauthorblockA{\IEEEauthorrefmark{1}University of Southern California, CA 90089, USA. Email: \{yangcai, molisch\}@usc.edu}
	\IEEEauthorblockA{\IEEEauthorrefmark{2}New York University, NY 10012, USA. Email: \{jllorca, atulino\}@nyu.edu}
	\IEEEauthorblockA{\IEEEauthorrefmark{3}University\`{a} degli Studi di Napoli Federico II, Naples 80138, Italy. Email: antoniamaria.tulino@unina.it}
	\thanks{
		An extended version of this paper is submitted to the IEEE Transactions on Communications \cite{cai2022delay_arxiv}.
	}
}


\maketitle

\IEEEpubid{
\begin{minipage}{2\columnwidth}
	\centering
	{\footnotesize
	\vspace{80pt}
	\copyright\ 2021 {IEEE}. Personal use of this material is permitted. Permission from {IEEE} must be obtained for all other uses, in any current or future media, including reprinting/republishing this material for advertising or promotional purposes, creating new collective works, for resale or redistribution to servers or lists, or reuse of any copyrighted component of this work in other works.
	\cite{cai2021delay}, DOI: \url{10.1109/ICC42927.2021.9500573}.
	}
\end{minipage}
}

\IEEEtitleabstractindextext{
\begin{abstract}
The timely delivery of resource-intensive and latency-sensitive services (e.g., industrial automation, augmented reality) over distributed computing networks (e.g., mobile edge computing) is drawing increasing attention. Motivated by the insufficiency of {\em average} delay performance guarantees provided by existing studies, we focus on the critical goal of delivering next generation real-time services ahead of corresponding deadlines {\em on a per-packet basis}, while minimizing overall cloud network resource cost. We introduce a novel queuing system that is able to track {\em data packets' lifetime} and formalize the optimal cloud network control problem with strict deadline constraints. After illustrating the main challenges in delivering packets to their destinations before getting dropped due to lifetime expiry, we construct an equivalent formulation, where relaxed flow conservation allows leveraging Lyapunov optimization to derive a provably near-optimal fully distributed algorithm for the original problem. Numerical results validate the theoretical analysis and show the superior performance of the proposed control policy compared with state-of-the-art cloud network control.
\end{abstract}}

\acresetall

\IEEEdisplaynontitleabstractindextext

\IEEEpeerreviewmaketitle


\section{Introduction}

The past decade has seen a proliferation of resource- and interaction-intensive applications, such as real-time computer vision, autonomous transportation, machine control in Industry 4.0, multiuser video conferencing, and augmented/virtual reality \cite{cai2022metaverse,mach17mecsurvey}, which we collectively refer to as \ac{agi} services.
In addition to the communication resources needed for the delivery of data streams to corresponding destinations, \ac{agi} services also require a significant amount of computation resources for the real-time processing of source data streams.
In contrast, \acp{ue} are evolving towards increasingly small, portable devices (and inevitably, with constrained power and computing capabilities), pushing the need to offload many computing tasks to the cloud, especially those running advanced architectures such as fog and \ac{mec}, which deploy computation resources closer to the end users in order to strike a better balance between access delay and resource efficiency.

Delay and cost are thus two essential metrics when evaluating the performance of AgI service delivery. From the consumers' perspective, excessive end-to-end delays can significantly impact quality of experience (QoE), especially for delay-sensitive AgI applications where packets must be delivered by a strict deadline in order to be effective.
In this context, {\em timely throughput}, which measures the rate of effective packet delivery (i.e., within-deadline packet delivery rate), becomes the appropriate performance metric \cite{CheHua:J18}.
In contrast, network operators care about the overall resource (e.g., computation, communication) consumption (and associated cost) needed to support the  dynamic service requests raised by end users,
which are dictated by the decision of route selection, function execution, and the corresponding resource allocation \cite{BarLloTulRam:C15}.

Previous studies have shown that the cloud network control problem \cite{FenLloTulMol:J18a,cai2020mec,cai2021multicast,cai2022multicast_arxiv,cai2022xpipelines_arxiv,cai2022CCC_arxiv}, involving packet routing and processing decisions over a distributed computing network, can be connected to the {\em packet routing} problem in traditional communication networks via a properly constructed {\em cloud-augmented} or {\em layered-graph} formulation \cite{ZhaSinLloTulMod:C18}. For packet routing, many dynamic control policies have been developed aimed at maximizing network throughput, including the celebrated \ac{bp} algorithm \cite{TasEph:J92} and its extension, the \ac{ldp} control approach \cite{Nee:B10} that, in addition, optimizes network resource cost (e.g., energy expenditure). 
While having the remarkable advantage of achieving throughput optimal performance via simple local policies without requiring any knowledge of network topology and traffic demands, both \ac{bp} and \ac{ldp} approaches can suffer from poor (average) delay performance, especially in low congestion scenarios, where packets can take unnecessary long and sometimes even cyclic paths.
In an attempt to address this problem, \cite{YinShaRedLiu:J11} proposed a combination \ac{bp} and shortest-path routing; while \cite{SinMod:J18} designed a centralized source routing approach, referred to as UMW, which reduces the average delay by dynamically selecting an acyclic route for each incoming packet, however requiring global network information.

Going beyond average delay and analyzing per-packet delay performance is a much more challenging problem with much fewer known results.
In \cite{Nee:C11}, a variant of the \ac{bp} algorithm is developed that provides worst-case delay guarantees by allowing packet drops, leading to a tradeoff between delay and achievable throughput; however, the relationship is not tight enough for practical purposes.
In \cite{singh2018delay}, the authors formulate the problem of timely throughput maximization as a \ac{cmdp}, and address it by solving the {\em single-packet routing} problem separately for each packet; yet, its computational complexity is prohibitive for practical implementation.
A more comprehensive literature review is presented in \cite{cai2022delay_arxiv}.

In this paper, we investigate the problem of multi-hop distributed cloud network control with the goal of delivering multiple \ac{agi} services with strict deadline constraints on a per-packet basis, while minimizing overall resource cost.
Our contributions can be summarized as follows:
\begin{enumerate}
	\item We characterize the {\em delay-constrained network capacity region}
	leveraging a novel {\em lifetime-driven} generalized flow conservation law.
	\item We develop a fully distributed control algorithm for the delivery of delay-sensitive services, shown to achieve {\em timely-throughput} optimality while minimizing overall resource cost.
	\item We present numerical results illustrating the delay-constrained network capacity region and the superior delay-cost performance tradeoff of the proposed policy.
\end{enumerate}

\section{System Model}\label{sec:system_model}

Due to space limitations,
in this paper we describe the proposed approach for the {\em packet routing} problem and refer the reader to the longer version in \cite{cai2022delay_arxiv} for its cloud network control generalization.\footnote{
	The generalization is based on the layered-graph technique \cite{ZhaSinLloTulMod:C18}.
}
Nonetheless, we do use the general cloud network control model for the numerical results in Section \ref{sec:experiments}.

Consider the packet routing network modeled by a directed graph $\Set{G} = (\Set{V}, \Set{E})$, with $\Set{V}$ and $\Set{E}$ denoting the node and edge set, respectively.
The nodes can transmit packets via the links $(i, j) \in \Set{E}$ between them, and we denote by $\delta_i^-$ and $\delta_i^+$ the incoming and outgoing sets of node $i$, respectively.

Time is slotted, and we quantify the available communication resource and the associated cost as \cite{FenLloTulMol:J18a}
\begin{itemize}
  \item $C_{ij}$: the average transmission capacity,\footnote{
  As an extension to this work, the problem formulated under a more realistic scenario with {\em peak} link capacity constraint is also studied in \cite{cai2022delay_arxiv}.
  The solution closely follows the principle of the methodology presented in this paper.
  }
  i.e., the maximum average transmission rate of link $(i, j)$;
  \item $e_{ij}$: the transmission cost, i.e., the cost of transmitting one unit size of data, on link $(i, j)$. 
\end{itemize}

The problem of interest is to deliver the packets to their destinations ahead of corresponding deadlines, to which {\em data packets' lifetime} is closely related, defined as the number of time slots before the information contained in the packet becomes useless.
A packet is called {\em effective} if its remaining lifetime $l > 0$, and {\em outdated} otherwise.
We assume that only the delivery of effective packets counts (i.e., outdatedness is as bad as packet loss); the associated metric, {\em timely throughput}, i.e., the rate of effective packets delivery, is employed to characterize the capability of the communications network.

\subsection{Request Model}

For ease of exposition, we formulate the problem for a single destination-based commodity described by a destination node (or user) $d \in \Set{V}$ that requests of a given application, with the straightforward extension to multiple commodities given in  \cite{FenLloTulMol:J18a}.
We assume that the input packets of the given commodity can originate at any node $i\in \Set{V} \setminus \{d\}$ (restriction of a specific source node set is straightforward) and that the source of each packet is aware of its  lifetime $l\in \Set{L} \triangleq \{1, \cdots, L\}$ at birth, where $L$ denotes the maximal lifetime.
We denote by $a_i^{(l)}(t)$ the number of lifetime-$l$ packets arriving at node $i$ on time slot $t$, which is assumed to be i.i.d. over time (with an upper bound of $A_{\max}$);
besides, the mean arrival rate is defined as $\lambda_i^{(l)} \triangleq \mathbb{E}\big\{a_i^{(l)}(t)\big\}$, and the collection $\V{\lambda} = \big\{ \lambda_i^{(l)}: \forall \, i\in \Set{V}, l \in \Set{L} \big\}$ is called the arrival vector.

\subsection{Queuing System}

We construct a queuing system $\V{Q}(t)$ that includes distinct queues for packets of different {\em current lifetimes} $l \in \Set{L}$, 
and denote by $Q_i^{(l)}(t)$ the queue backlog (i.e., number of packets in the queue) of lifetime $l$ packets at node $i$ on time slot $t$.

Each time slot is divided into two phases.
In the {\em decision} phase, the nodes make and accomplish the transmission decisions (which packets are sent out, to which neighboring node);
in the {\em receiving} phase, the incoming packets, including those from neighboring nodes and the exogenous packets, are collected and loaded into the queuing system.
Let $x_{ij}^{(l)}(t)$ be the number of lifetime $l$ packets that are sent from node $i$ to $j$ in time slot $t$; we refer to it as {\em flow variable}.

In general, the queuing dynamics are given by\footnote{
  The lifetime of the exogenous packets $a_i^{(l)}(t)$ are counted starting from the beginning of next time slot, i.e., the point when they are available for use; the lifetime $l$ of the transmitted packets $x_{ij}^{(l)}$ is the current lifetime, which are available at node $j$ in next time slot with a lifetime of $l-1$. 
}
\begin{align}\label{eq:queue_dynamics_1}
\hspace{-.12cm} Q_i^{(l)}(t+1) = Q_i^{(l+1)}(t) - x_{i\to}^{(l+1)}(t) + x_{\to i}^{(l+1)}(t) + a_i^{(l)}(t)
\end{align}
where $x_{\to i}^{(l)}(t) = \sum_{j\in \delta_i^-} x_{ji}^{(l)}(t)$ and $x_{i \to}^{(l)}(t) = \sum_{j\in \delta_i^+} x_{ij}^{(l)}(t)$ denote the total incoming and outgoing packets of node $i$.

In addition, we make the following assumptions: 1) outdated packets (not contribute to timely throughput) are dropped, i.e.,
\begin{align}\label{eq:queue_dynamics_2}
Q_i^{(0)}(t) = 0, \quad \forall\, i \in \Set{V},
\end{align}
and 2) for the destination node $d$, any effective packet is consumed as soon as it arrives, and therefore
\begin{align}\label{eq:queue_dynamics_3}
Q_d^{(l)}(t) = 0, \quad \forall\, l \in \Set{L}.
\end{align}

\subsection{Admissible Policy Space}\label{sec:admissible}

The considered control policies make decisions on packet routing and scheduling in each time slot,\footnote{
	Note that a packet can be discarded in network due to lifetime expiry.
	In this sense, the designed policy also deals with admission control aspect.
}
which are specified by the flow variables $\V{x}(t) = \big\{ x_{ij}^{(l)}(t): \forall\, (i, j)\in \Set{E}, l \in \Set{L} \big\}$.
We restrict to the space of {\em admissible} control policies with the decided flow variables satisfying the following conditions:
\begin{enumerate}
	\item non-negativity constraint, i.e.,
	\begin{align}
	x_{ij}^{(l)}(t) \geq 0\text{ for }\forall\,(i, j)\in \Set{E},\text{ or }\V{x}(t) \succeq 0;
	\end{align}
	\item average capacity constraint, i.e.,
	\begin{align}\label{eq:capacity}
	\avg{\E{x_{ij}(t)}} \leq C_{ij},\ \forall (i, j)\in \Set{E}
	\end{align}
	where $x_{ij}(t)\triangleq \sum_{l\in \Set{L}} x_{ij}^{(l)}(t) $, and $\avg{ z(t) } $ denotes the expected long-term average operation, defined as $\avg{ z(t) } = \lim_{T\to\infty} \frac{1}{T} \sum_{t=0}^{T-1}{ z(t) }$;
	\item availability constraint, which requires the number of scheduled outgoing packets not to exceed those in the current queuing system, i.e.,
	\begin{align}\label{eq:availability}
	x_{i \to}^{(l)}(t) \leq Q_i^{(l)}(t),\ \forall\,i\in \Set{V},\, l\in \Set{L};
	\end{align}
	\item reliability constraint, i.e.,
	\begin{align}\label{eq:reliability}
	\hspace{-2pt} \avg{ \E{ x_{\to d}(t) } } \triangleq \sum\nolimits_{l \in \Set{L}} \avg{ \E{x_{\to d}^{(l)}(t)} } 
	\geq \gamma \| \V{\lambda} \|_1
	\end{align}
	where $\gamma$ is named the {\em reliability level}, and $\| \V{\lambda} \|_1$ is the total arrival rate of the application.
\end{enumerate}

The reliability constraint quantifies the extent to which the considered application can tolerate packet loss.
It implies that a percentage of up to $(1-\gamma)$ of the incoming packets can be dropped without causing a significant performance loss.

The instantaneous cost of the decision $\V{x}(t)$ is given by
\begin{align}\label{eq:cost}
h(t) = h(\V{x}(t)) = \sum\nolimits_{(i, j)\in \Set{E}} e_{ij} x_{ij}(t) = \langle \V{e}, \V{x}(t) \rangle
\end{align}
where $\langle \cdot, \cdot \rangle$ denotes the inner product of the two vectors.

\begin{remark}\label{remark:q}
In the existing literature of stochastic network optimization, the {\em assigned} flow has gained widespread use (e.g., \cite{FenLloTulMol:J18a,Nee:B10,Nee:C11,TasEph:J92}), which is different from the actual flow in that it does not take the availability constraint \eqref{eq:availability} into account, and thus the decision space in each time slot does not depend on the current queuing status.
{\em Dummy} packets are created when there are not sufficient packets in the queue to support the decision.
The formulation is not suitable for the considered problem, because the reliability constraint \eqref{eq:reliability} is imposed on {\em actual} packets received by the destination node $d$;
while in the previous formulation, the flow variables $x_{jd}^{(l)}(t)\ (j \in \delta_d^-)$ can include dummy packets.
\end{remark}

\subsection{Problem Formulation}

The goal of this work is to develop an admissible control policy that guarantees reliable packet delivery, while minimizing network resource cost.
Formally, we aim to find the policy that makes decisions $\{ \V{x}(t): t\geq 0 \}$ satisfying
\begin{subequations}\label{eq:p1}
\begin{align}
\mathscr{P}_1: \ & \min_{\V{x}(t) \succeq 0}\ \avg{ \E{ h(\V{x}(t)) } } \\
& \ \st\ \  \avg{ \E{ x_{\to d}(t) } } \geq \gamma \| \V{\lambda} \|_1 \label{eq:p1_reliability} \\
& \hspace{.42in} \avg{\E{x_{ij}(t)}} \leq C_{ij},\ \forall \, (i,j) \in \Set{E} \label{eq:p1_capacity} \\
& \hspace{.42in} x_{i\to}^{(l)}(t) \leq Q_i^{(l)}(t),\ \forall \, i\in \Set{V}, l \in \Set{L} \label{eq:p1_availability} \\
& \hspace{.42in} \V{Q}(t) \text{ evolves by \eqref{eq:queue_dynamics_1} -- \eqref{eq:queue_dynamics_3}}. \label{eq:p1_queue}
\end{align}
\end{subequations}

We emphasize that the above problem cannot be addressed by the {\em \ac{ldp}} approach, because (i) the queuing system \eqref{eq:p1_queue} allows packet drops by \eqref{eq:queue_dynamics_2}, and (ii) we employ the actual flow as the decision variable, i.e., constraint \eqref{eq:p1_availability}, which are different from the standard formulation.

In general, \eqref{eq:p1} can be interpreted as a \ac{cmdp} problem \cite{singh2018delay}.
However, we emphasize that the related state vector $\V{Q}(t)$ and action vector $\V{x}(t)$ are {\em network-wide}, which leads to a dramatically increasing state-action space, and the complexity of the standard solution \cite{Alt:B99} is prohibitive for practical implementation.

\section{The Equivalent Problem} \label{sec:approximate}

In this section, we present a new problem $\mathscr{P}_2$, which is referred to as the {\em virtual network}.
We prove the equivalence of the two problems in terms of flow space and capacity region, and use $\mathscr{P}_2$ as a stepping-stone to find the solution to $\mathscr{P}_1$.

\subsection{The Equivalent Problem}\label{sec:relax}

The new problem is cast as
\begin{subequations}\label{eq:problem_2}
\begin{align}
\hspace{-.1in}\mathscr{P}_2: \ & \min_{\V{x}(t)\succeq 0} \ \avg{ \E{ h(\V{x}(t)) } } \\
& \ \st\ \ \avg{ \E{ x_{\to d}(t) } } \geq \gamma \| \V{\lambda} \|_1 \label{eq:p2_reliability} \\
& \hspace{.42in} x_{ij}(t) \leq C_{ij} \label{eq:p2_capacity} \\
& \hspace{.1in} \avg{ \E{ x_{i\to}^{(\geq l)}(t) } } \leq \avg{ \E{x_{\to i}^{(\geq l+1)}(t)} } + \lambda_i^{(\geq l)} \label{eq:p2_causality}
\end{align}
\end{subequations}
where the superscript $(\geq l)$ indicates that the term includes all the lifetimes $\ell$ satisfying $\ell \geq l$, e.g., $x_{ij}^{(\geq l)}(t) = \sum_{\ell = l}^{L}{ x_{ij}^{(\ell)}(t) }$.

\subsubsection{Virtual Queues}

The crucial difference in deriving $\mathscr{P}_2$ is to eliminate (i) the unconventional queuing system \eqref{eq:p1_queue} and (ii) \eqref{eq:p1_availability} that makes $\V{x}(t)$ dependent on $\V{Q}(t)$, i.e., the two factors prohibiting direct application of the \ac{ldp} approach.
Instead, we introduce the relaxed {\em causality} constraint \eqref{eq:p2_causality} (see Proposition \ref{prop:causality_constraint}) to state the fact that the lifetime of the packets must decrease as they traverse any node $i$.

Although there is no explicit queuing system involved in $\mathscr{P}_2$, it consists of long-term average objective and constraints, which can be addressed by the \ac{ldp} approach via the use of virtual queues \cite{Nee:B10} (we denote the solution by $\V{\nu}(t)$).
More concretely, ensuring constraints \eqref{eq:p2_reliability} and \eqref{eq:p2_causality} is equivalent to stabilize the following virtual queues:
\begin{subequations}
\begin{align}
U_d(t+1) & = \max\big\{ 0,\, U_d(t) + \gamma A(t) - \nu_{\to d}(t) \big\} \label{eq:virtual_sink} \\
U_i^{(l)}(t+1) & = \max \big\{0,\, U_i^{(l)}(t) + \nu_{i\to}^{(\geq l)}(t) - \nu_{\to i}^{(\geq l+1)}(t) \nonumber \\
& \hspace{30 pt} - a_i^{(\geq l)}(t) \big\}\quad (i\in \Set{V}\setminus \{d\}, l\in \Set{L}) \label{eq:virtual_intermediate}
\end{align}
\end{subequations}
where $A(t) = \sum_{i\in \Set{V}} \sum_{l\in \Set{L}} a_i^{(l)}(t)$ is the total amount of packets arriving at the network in time slot $t$.\footnote{
  Here we use $A(t)$ instead of $\| \V{\lambda} \|_1$ as the latter information is usually not available in practice;
  furthermore, if the arrival information cannot be obtained immediately, delayed information, i.e., $A(t-\tau)$ with $\tau > 0$, can be used as an alternative, which does not impact the result of time average.
}
We refer to \eqref{eq:virtual_sink} and \eqref{eq:virtual_intermediate} as the virtual queues at node $d$ and $i$, respectively.

\subsubsection{Physical Interpretations}

$\mathscr{P}_2$ describes a virtual network modeling each node as a {\em data-reservoir}, which has access to abundant (virtual) packets of any lifetime.
As neighboring nodes request packets from node $i$, it supplies the needs by using the virtual packets from the reservoir in advance, which are compensated when the node receives incoming packets of the same lifetime.
The virtual queues can be roughly explained as the {\em accumulated data deficits} (outgoing flow minus incoming flow) of the corresponding data-reservoirs;
specially, in \eqref{eq:virtual_sink}, the destination reservoir {\em sends out} $\gamma A(t)$ packets to the end user (as is required by the reliability constraint), while {\em receiving} $\nu_{\to d}(t)$ in return.
When \eqref{eq:p2_causality} is satisfied, node $i$ no longer embezzles the virtual packets from its reservoir;
and if it is true for all nodes, the data streams in the network include only actual packets.
The resulting flow assignment (defined in next subsection) can instruct the packets to find their paths in the actual network.

\subsection{Relationships Between $\mathscr{P}_1$ and $\mathscr{P}_2$} \label{sec:relation}

Similar to Section \ref{sec:admissible}, for a given pair of $(\V{\lambda}, \gamma)$, we define a policy $p$ to be {\em admissible} for $\mathscr{P}_2$ if it satisfies \eqref{eq:p2_reliability} -- \eqref{eq:p2_causality}.
In addition, suppose an admissible policy $p\in \Set{A}_n\ (n=1,2)$ makes decisions $\V{x}_p(t) = \big\{ x_{ij}^{(l)}(t) : (i, j)\in \Set{E},\, l\in \Set{L}, \, t\geq 0 \big\}$,
then the associated {\em flow assignment} is defined as $\V{x}_p = \avg{ \E{ \V{x}_p(t) } }$, which collects the transmission rates of all links.

\begin{definition}
For given $(\V{\lambda}, \gamma)$, the {\em admissible policy space} $\Set{A}_n$ is defined as the collection of all admissible control policies for problem $\mathscr{P}_n\ (n=1,2)$.
\end{definition}

\begin{definition}
The {\em network capacity region} $\Lambda_n$ is defined as the set of $(\V{\lambda}, \gamma)$ pairs, under which the admissible policy space $\Set{A}_n$ is non-empty $(n = 1, 2)$.
\end{definition}

\begin{definition}
For given $(\V{\lambda}, \gamma) \in \Lambda_n$, the {\em flow space} is defined as the set of all flow assignments that can be achieved by the admissible policies, i.e., $\Gamma_n = \big\{ \V{x}_p: p\in \Set{A}_n \big\}\ (n = 1, 2)$.
\end{definition}

Next, we present the relationships between the two problems, in terms of the above quantities.

\begin{proposition}\label{prop:causality_constraint}
\eqref{eq:p1_availability} implies \eqref{eq:p2_causality}, and \eqref{eq:p2_capacity} implies \eqref{eq:p1_capacity}.
\end{proposition}

\begin{proposition}\label{thm:cap_region}
For a given network, the capacity regions of the two problems are identical, i.e., $\Lambda_1 = \Lambda_2$.

A pair $(\V{\lambda},\gamma)$ is within the capacity region $\Lambda_n\ (n = 1,2)$ if and only if there exist flow variables $\V{x} = \{ x^{(l)}_{ij} \geq 0: \forall\, (i,j)\in \Set{E}, l \in \Set{L} \}$,
such that $\forall\, i\in \Set{V}$, $(i,j)\in \Set{E}$, $l \in \Set{L}$, 
\begin{subequations}\label{eq:capacity_region}
\begin{align}
x_{\to d} & \geq \gamma \| \V{\lambda} \|_1 \label{eq:cr_reliability} \\
x_{ij} & \leq C_{ij},\ \forall\,(i,j)\in \Set{E} \label{eq:cr_capacity} \\
x_{\to i}^{(\geq l+1)} + \lambda_i^{(\geq l)} & \geq x_{i\to}^{(\geq l)},\ \forall\,i\in \Set{V}, l\in \Set{L} \label{eq:flow_conserve} \\
x_{ij}^{(0)} = x_{dk}^{(l)} & = 0,\ \forall\,k\in \delta_d^{+},(i, j)\in \Set{E}, l\in \Set{L}.
\end{align}
\end{subequations}

Furthermore, for any point within the capacity region, there exists a randomized policy $*$ to support it while attaining optimal cost performance. 
\end{proposition}

\begin{proposition}\label{prop:flow_space}
For any point $(\V{\lambda}, \gamma) \in \Lambda_1 = \Lambda_2$, the associated flow space $\Gamma_1 = \Gamma_2$.
\end{proposition}

\begin{IEEEproof}
All proofs can be found in \cite{cai2022delay_arxiv}.
In Proposition \ref{thm:cap_region}, given $\V{x}$ satisfying \eqref{eq:capacity_region}, the feasible randomized policy $*$ (for $\mathscr{P}_1$) operates as follows:
in each time slot, any packet of lifetime $l\in \Set{L}$ at node $i\in \Set{V}$ has a probability
\begin{align} \label{eq:p2_randomize}
\alpha_i^{(l)}(j) = x_{ij}^{(l)} \Big/ \Big( x_{\to i}^{(\geq l+1)} + \lambda_i^{(\geq l)} - x_{i\to}^{(\geq l+1)} \Big)
\end{align}
to be sent to node $j$, and stay in node $i$ otherwise; and
the policy $*$ achieves the flow assignment $\V{x}$.
\end{IEEEproof}

The previous propositions can be explained as follows:
by Proposition \ref{prop:causality_constraint}, in general, the admissible policy spaces $\Set{A}_1 \nsubseteq \Set{A}_2$ and $\Set{A}_2 \nsubseteq \Set{A}_1$;
Proposition \ref{thm:cap_region} suggests that they lead to the same capacity regions by presenting an explicit identical characterization \eqref{eq:capacity_region}, where \eqref{eq:flow_conserve} is interpreted as the generalized {\em flow conservation} law when considering the packets' lifetime;
Proposition \ref{prop:flow_space} further shows that $\mathscr{P}_1$ and $\mathscr{P}_2$ share the same flow space for any $(\V{\lambda}, \gamma)$, which is a crucial property for the considered problem, where the two metrics of interest, i.e., timely throughput \eqref{eq:reliability} and resource cost \eqref{eq:cost}, are both {\em linear} functions of the flow assignment.

\begin{corollary}\label{coro:opt_value}
$\mathscr{P}_1$ and $\mathscr{P}_2$ have the same optimal value.
\end{corollary}

\begin{IEEEproof}
Because they have the same flow space.
\end{IEEEproof}

\section{Proposed Control Policy} \label{sec:average_solution}

In this section, we provide a solution for $\mathscr{P}_2$ leveraging Lyapunov optimization theory, and take advantage of Propositions \ref{thm:cap_region} and \ref{prop:flow_space} to develop an algorithm for $\mathscr{P}_1$ based on it.

\subsection{Solution to the Virtual Network Problem} \label{sec:ldp_alg}

We define the Lyapunov function as $L(t) = \|\V{U}(t)\|_2^2 / 2$, and Lyapunov drift $\Delta(\V{U}(t)) = L(t+1) - L(t)$.
The \ac{ldp} approach advocates to minimize a linear combination of the Lyapunov drift (see \cite{cai2022delay_arxiv}) and the cost function weighted by a tunable parameter $V$ (which controls the tradeoff between network congestion and operational cost), i.e.,
\begin{align}\label{eq:virtual_ub}
\Delta( \V{U}(t) ) + V h( \V{\nu}(t) )
\leq B - \langle \tilde{\V{a}}, \V{U}(t) \rangle - \langle \V{w}(t), \V{\nu}(t) \rangle
\end{align}
where $\tilde{\V{a}} = \big\{ a_d(t) - \gamma A(t) \big\} \cup \big\{ a_i^{(\geq l)}(t) : \forall\,i\in \Set{V}\setminus\{d\}, l\in \Set{L} \big\}$, $B$ is a constant, and the weights $\V{w}(t)$ are given by
\begin{align}\label{eq:weight}
w_{ij}^{(l)}(t) = - V e_{ij} - U_i^{(\leq l)}(t) +
\begin{cases}
U_d(t) & j = d \\
U_j^{(\leq l-1)}(t) & j\ne d
\end{cases}
\end{align}
where the superscript $^{(\leq l)}$ refers to the operation of $\sum_{\ell = 1}^{l}$.

To sum up, the developed algorithm aims to solve the following problem in each time slot
\begin{subequations}\label{eq:virtual_opt}
\begin{align}
\max_{\V{\nu}(t) \succeq 0} \ \langle \V{w}(t), \V{\nu}(t) \rangle,\ \st \nu_{ij}(t) \leq C_{ij},\ \forall\, (i,j)\in \Set{E}.
\end{align}
\end{subequations}
The solution is in the {\em max-weight} fashion.
More concretely, for each link $(i, j)$, we first find the lifetime $l^\star$ with the largest weight, and spend all the transmission resource to transmit packets of this lifetime if the associated weight is positive.
To sum up, the optimal flow assignment is
\begin{align}
\nu_{ij}^{(l)}(t) = C_{ij}\,\mathbb{I}\big\{ l = l^\star, w_{ij}^{(l^\star)}(t) > 0 \big\}
\end{align}
where the optimal lifetime choice is $l^\star = \argmax_{l\in \Set{L}}\, w_{ij}^{(l)}(t)$, and $\mathbb{I}\{ \cdot \}$ denotes the indicator function, which equals $1$ when the two events in the bracket are both true.

Due to the additive form of the objective function, which is composed of sub-problems that can be completed in each individual node, the algorithm can be implemented in a fully distributed manner.

\subsection{Performance Analysis}

In this part, we present a proposition analyzing the performance of the proposed control policy related to the timely throughput (for reliability constraint \eqref{eq:reliability}) and resource cost.

\begin{definition}[$\varepsilon$-Convergence Time]
The $\varepsilon$-convergence time $t_\varepsilon$ is the first time index, such that the achieved reliability level is within a gap of $\varepsilon$ from the desired value ever after, i.e.,
\begin{align}
t_\varepsilon \triangleq \min_\tau \Big\{ \sup_{s\geq \tau} \Big[\gamma \| \V{\lambda} \|_1 -  \sum_{t=0}^{s-1} \frac{\E{ \nu_{\to d}(t) }}{s} \Big] \leq \varepsilon \Big\}.
\end{align}
\end{definition}

\begin{proposition}\label{prop:cost_V}
For any point in the interior of the capacity region, under the proposed algorithm, the virtual queues are mean rate stable  with a convergence time $t_\varepsilon \sim \mathcal{O}(V)$ for any $\varepsilon > 0$, and the achieved cost performance satisfies
\begin{align} \label{eq:achieved_cost}
\avg{ \E{ h_2(\V{\nu}(t)) } } \leq h_2^\star( \V{\lambda}, \gamma ) + B/V
\end{align}
where $h_2^\star( \V{\lambda}, \gamma )$ denotes the optimal cost performance that can be achieved under $( \V{\lambda}, \gamma )$ in $\mathscr{P}_2$.
\end{proposition}

\begin{IEEEproof}
See \cite{cai2022delay_arxiv}.
\end{IEEEproof}

From the above proposition, we find that by pushing the parameter $V\to \infty$, the achieved cost performance approaches the optimal cost (since the gap $B/V$ vanishes), while compromising the convergence time.

\subsection{Flow Matching}

In this section, we develop a near-optimal control policy for $\mathscr{P}_1$.
According to Proposition \ref{prop:flow_space} and Corollary \ref{coro:opt_value}, there exists a randomized policy (specified by probability values $\V{\alpha}$) to be optimal, and we aim to find the solution to $\mathscr{P}_1$ in this categoty.
Instead of find $\V{\alpha}$ directly (not straightforward), the proposed approach will leverage \eqref{eq:p2_randomize} and calculate the parameters therein using empirical virtual flow decisions.
In the following, we denote the decided flow for $\mathscr{P}_1$ on time slot $t$ by $\V{\mu}(t) = \{ \mu_{ij}^{(l)}(t) \}$ (to distinguish it from $\V{\nu}(t)$).

The goal of the designed policy is to conform with the constraints in $\mathscr{P}_1$ (i.e., satisfying \eqref{eq:p1_reliability} -- \eqref{eq:p1_availability}), while pursuing the goal of {\em flow matching}, i.e., $\avg{ \V{\mu}(t) } = \avg{ \V{\nu}(t) }$.
The reason to set the above goal is two-fold.
(i) It ensures that the two algorithms attain the same throughput and cost performance (as mentioned earlier, both metrics are linear functions of the flow assignment); therefore, $\{\V{\mu}(t)\}$ satisfies the reliability constraint and achieves the same (and thus near-optimal by Corollary \ref{coro:opt_value}) cost performance as $\{\V{\nu}(t)\}$.
(ii) The existence of the policy is guaranteed (as a result of identical flow spaces).
Actually, given a feasible flow assignment $\V{x}$ satisfying \eqref{eq:capacity_region} (specifically, $\avg{ \V{\nu}(t) }$),  we are already aware of the construction procedure of the randomized policy $*$ to achieve it (see {\em Proof} to Proposition \ref{thm:cap_region} in Section \ref{sec:relation}).

While we do not wait until the exact value of $\avg{ \V{\nu}(t) }$ is obtained (actually it takes forever) to construct the randomized policy $*$\,; as an alternative, its empirical values are employed.
In each time slot, we first calculate the probability values $\V{\alpha}(t)$ according to \eqref{eq:p2_randomize}, using the {\em finite-horizon average} $\bar{\V{\nu}}(t) = \frac{1}{t} \sum_{\tau = 0}^{t-1}{ \V{\nu}(\tau) }$ as the flow assignment $\V{x}$, and estimating $\V{\lambda}$ from the empirical arrivals by $\hat{\lambda}_i^{(\geq l)} = \frac{1}{t} \sum_{\tau = 0}^{t-1}{ a_i^{(\geq l)}(\tau) }$; then node $i$ transmits packets of lifetime $l$ to node $j$ according to the obtained distribution.
It leads to a time-varying randomized policy, but we stress that $\bar{\V{\nu}}(t)$ converges to $\avg{ \V{\nu}(t) }$ asymptotically, which no longer changes over time.\footnote{
  It is possible that the finite-horizon average $\bar{\V{\nu}}(t)$ can violate \eqref{eq:flow_conserve} at some time slot, and thus does not make a qualified candidate for $\V{x}$.
  However, as is mentioned, $\lim_{t\to\infty}\bar{\V{\nu}}(t) = \avg{\V{\nu}(t)}$, which satisfies the constraints.
  With this asymptotic guarantee, when such violation occurs, we can choose not the update the control policy in that time slot.
}

In addition to deciding the virtual flow by the algorithm in Section \ref{sec:ldp_alg}, the developed randomized policy requires each node to record {\em its own} incoming and outgoing flows to calculate the probability values by \eqref{eq:p2_randomize}, which can be completed locally.
Therefore, the proposed design can operate in a fully distributed manner.

\begin{proposition} \label{prop:flow_matching}
For any point in the interior of the capacity region, the proposed control policy is admissible, while achieving the near-optimal cost performance of $h( \avg{\V{\nu}(t) } )$.
\end{proposition}

\begin{IEEEproof}
See \cite{cai2022delay_arxiv}.
\end{IEEEproof}

\section{Numerical Experiments} \label{sec:experiments}

In this section, we carry out several numerical experiments to evaluate the performance of the proposed design, based on the Abilene US continental network in Fig. \ref{fig:network}.

We take a simple \ac{agi} service for example,\footnote{
	In the setup of the experiments, we adopt a simple example for illustrative purposes, and we refer the readers to \cite{cai2022delay_arxiv} for more realistic \ac{agi} services.
}
which requires the source data-stream to be processed by one function, and we assume that each node (representing a data center) in the network can host the given service function.
To describe the function's computation resource requirement, we assume that a CPU (a measure of computing resource) is capable of processing the incoming data-stream at a rate of $50$ Mbps, and the output data-stream has the same size as the input.

The available network resource and the associated cost is described in the following: each node in the network has a an average resource consumption budget of $2$ CPUs and the associated cost is $1\,/$CPU at $i \in \{5, 6\}$, and $2\,/$CPU at other data centers; each link exhibits the same average transmission rate of $1$ Gbps, with a cost of $1\,/$Gb.

There are two clients, i.e., $(\text{source},\,\text{destination})$ pairs, requesting the service, i.e., $(1, 9)$ and $(3, 11)$, both at the reliability level of $\gamma = 90\%$.
The packets arrive at the source nodes according to independent Poisson processes of parameter $\lambda$.
The lifetime of all packets at birth equals to the maximum lifetime $L$ ($\geq 5$, including $4$ time slots for transmission via the shortest path, and $1$ time slot for processing).

\subsection{Network Capacity Region}

\begin{figure*}[t]
	\centering
	\begin{minipage}[t]{0.18\textwidth}
		\centering
		\vspace{-72pt}
		\includegraphics[width=\textwidth]{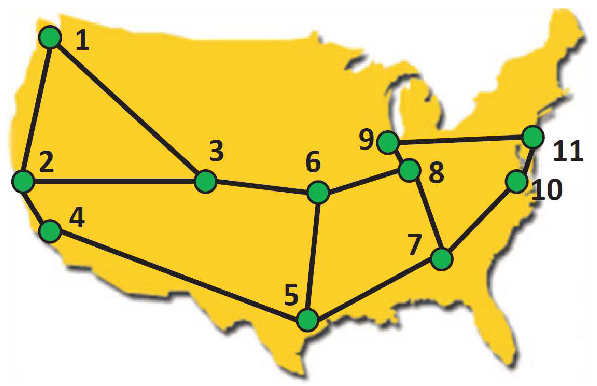}
		\vspace{0pt}
		\caption{The studied network.}
		\label{fig:network}
	\end{minipage}
	\hfill
	\begin{minipage}[t]{0.25\textwidth}
		\centering
		\includegraphics[width=\textwidth]{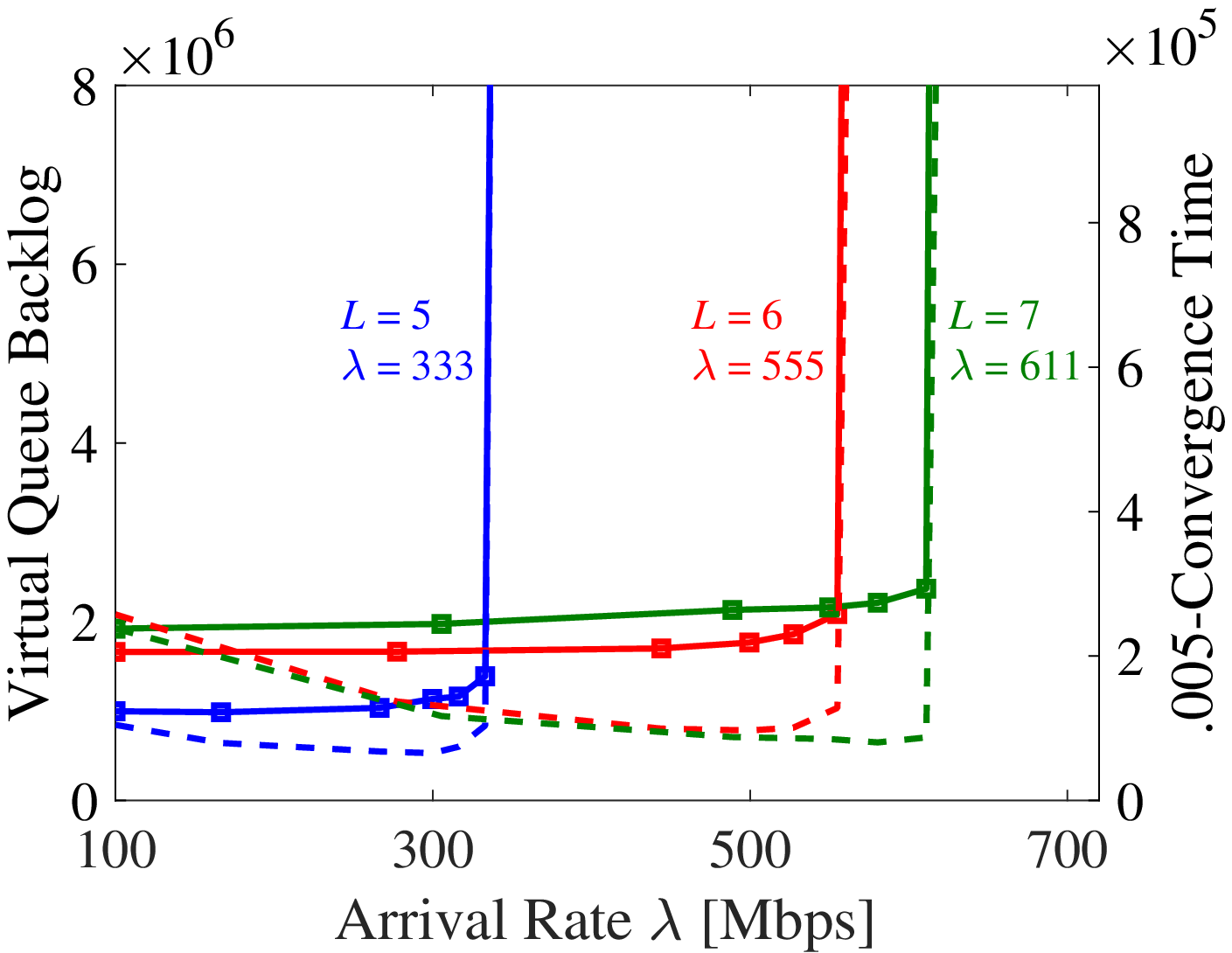}
		\vspace{-15pt}
		\caption{Capacity regions.}
		\label{fig:capacity_region}
	\end{minipage}
	\hfill
	\begin{minipage}[t]{0.25\textwidth}
		\centering
		\includegraphics[width=\textwidth]{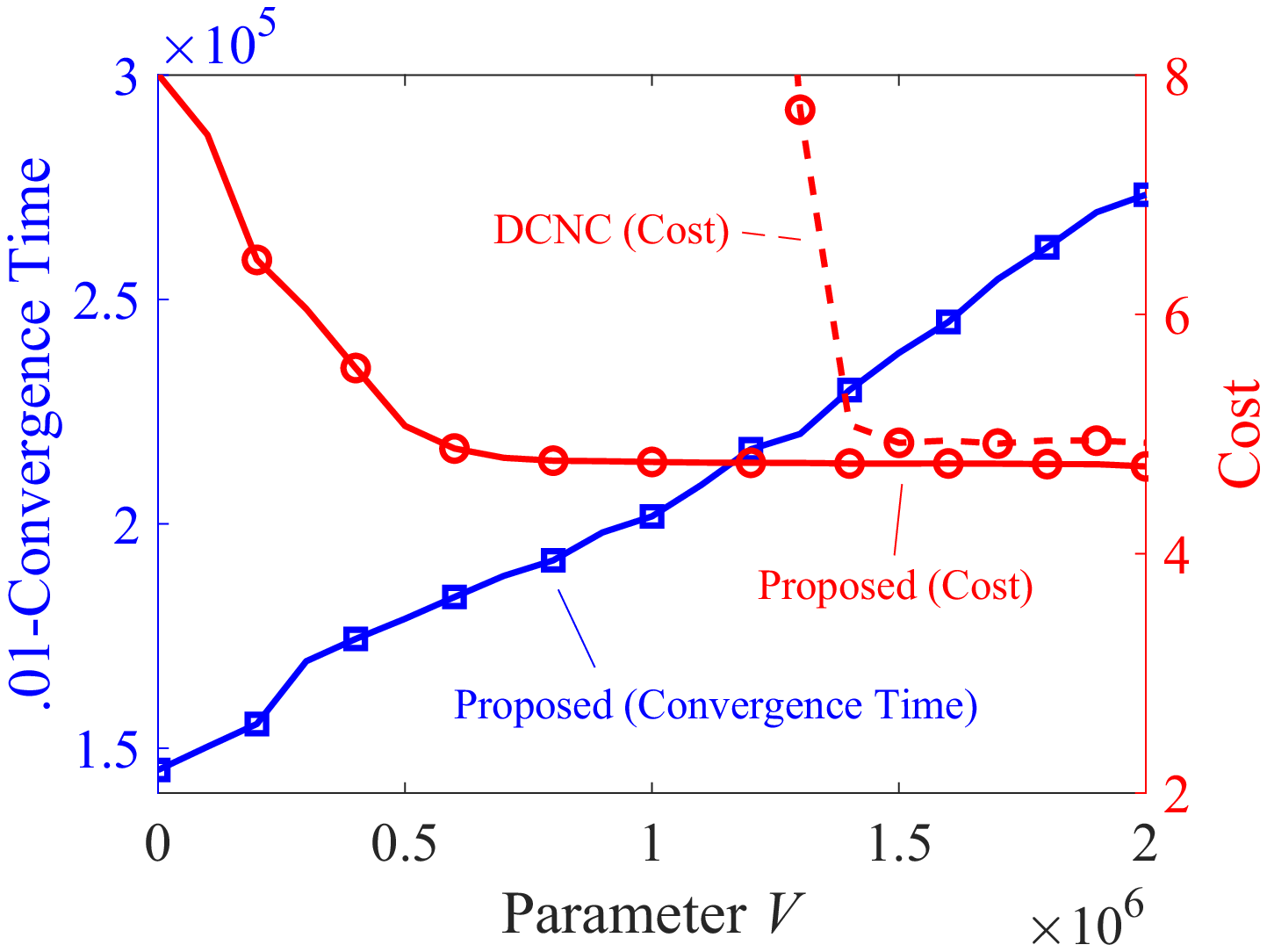}
		\vspace{-15pt}
		\caption{Tradeoff controlled by $V$.}
		\label{fig:V}
	\end{minipage}
	\hfill
	\begin{minipage}[t]{0.25\textwidth}
		\centering
		\includegraphics[width=\textwidth]{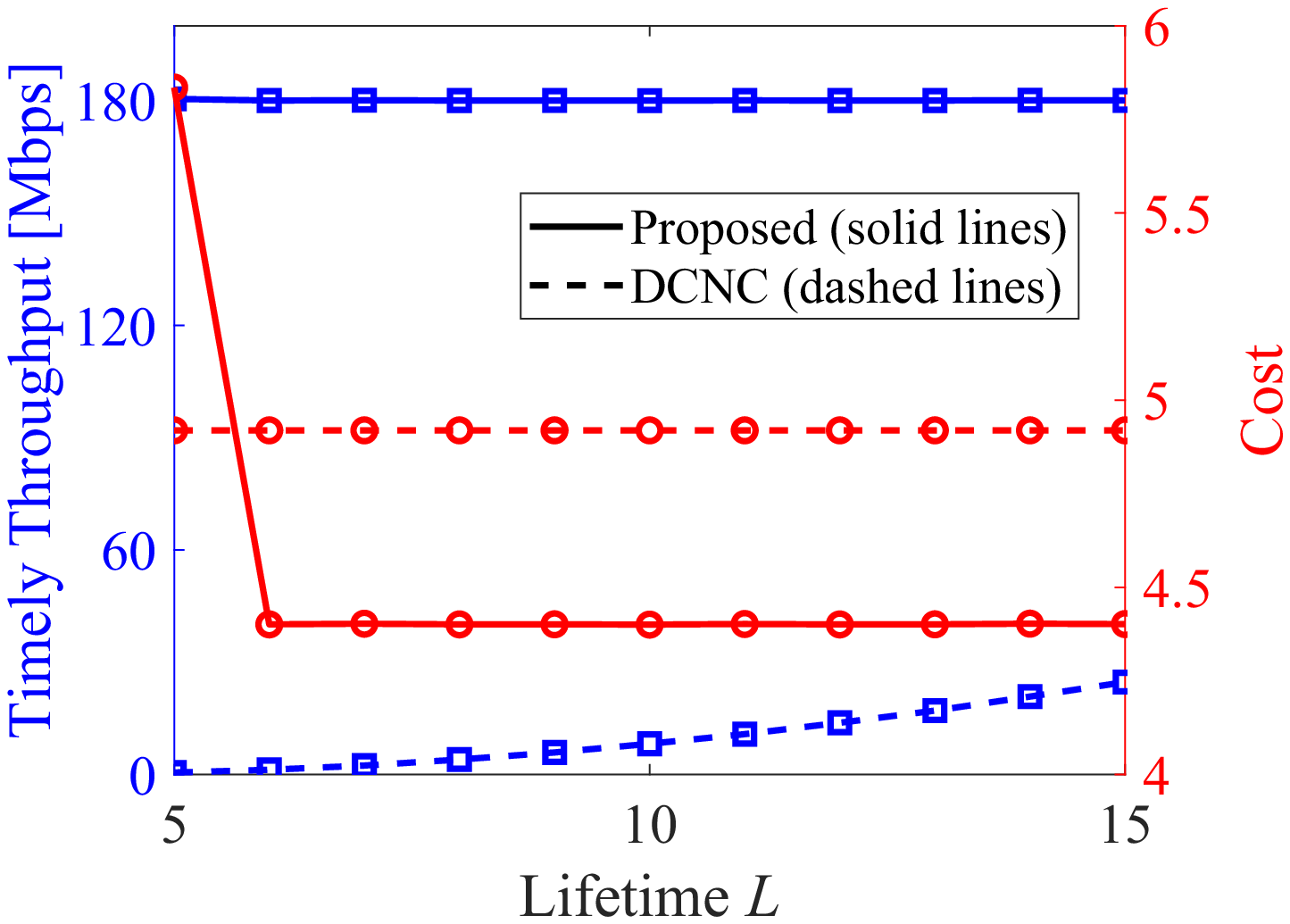}
		\vspace{-15pt}
		\caption{Effects of packets' lifetime.}
		\label{fig:L}
	\end{minipage}
	\vspace{-10pt}
\end{figure*}

We first study the network capacity regions achieved by the proposed algorithm, assuming different maximum lifetimes $L$.
We run the proposed algorithm (with $V = 0$) on the network for $1 \times 10^6$ time slots, recording the queue backlog of the virtual network, and the $0.005$-convergence time for the actual network (i.e., when achieved reliability level $\geq 89.5\%$).

The results are shown in Fig. \ref{fig:capacity_region}, and we make the following observations.
First, for fixed $L$, both the virtual queue (solid lines) and the convergence time (dashed lines) blow up after the arrival rate exceeds a critical point, which is interpreted as the boundary of the capacity region.
The result verifies that the virtual and the actual network have the same capacity regions (as stated in Proposition \ref{thm:cap_region}).
Second, by increasing the value of $L$, the capacity region enlarges, since the packets can detour to farther network locations for extra computing resources, while still arriving at the destinations within the deadline.
When $L = 5$, the packets must follow the shortest paths to the destinations, and the two clients must share the constrained computing resource at the common nodes on the paths (data centers $3, 6, 8, 9$);
when $L = 6$, client $(3, 11)$ can detour along $3\to 6\to 5\to 7\to 10\to 11$ for extra computing resource, boosting the capacity region;
when $L \geq 7$, the computing resources of the entire network are fully exploited.

\subsection{Effects of Parameter $V$}

Next, we study the convergence time and the resource cost achieved by the proposed algorithm under various $V$ (using $L = 7$ and $\lambda = 100$ Mbps). In addition, we compare the results with the  state-of-the-art min-cost max-throughput algorithm, referred to as DCNC \cite{FenLloTulMol:J18a}.

The result is depicted in Fig. \ref{fig:V}.
First, we focus on the performance of the proposed algorithm (solid lines), which exhibits the $[\mathcal{O}(V), \mathcal{O}(1/V)]$ tradeoff between the convergence time and the resource cost, as is presented in Proposition \ref{prop:cost_V}.
Second, for the DCNC algorithm, we observe from the experiment that the achieved timely throughput is around $10$ Mbps (i.e., a reliability level of $10\%$, see next section for more results), failing the reliability constraint (resulting in a convergence time of $\infty$).
For the cost performance, when $V \leq 1\times 10^6$, it leads to a much higher resource cost ($\geq 15$) than the proposed algorithm, since the packets can take cyclic paths to the destination, incurring extra cost; as $V$ grows, the cost reduces, while it is still higher than the proposed algorithm because it delivers all the packets (even the outdated ones).

\subsection{Effects of Lifetime $L$}

Finally, we present the timely throughput and the resource cost achieved by the proposed and the DCNC algorithms, under various maximum lifetimes ($\lambda = 100$ Mbps, and a large $V = 5\times 10^7$ is selected to ensure near-optimal resource cost).

As we can observe from Fig. \ref{fig:L}, the proposed algorithm attains a (sum) timely throughput of $180$ Mbps, i.e., a reliability level of $90\%$ for each client, where the reliability constraint holds with equality (the existence of the $.01$-convergence time in the previous experiment also supports the result).
In contrast, the DCNC algorithm achieves much lower timely throughput, e.g., $20$ Mbps when $L = 15$, where the packets are provided $10$ extra time slots for transmission.
Finally, we point out that the resource cost of the proposed algorithm significantly improves when $L$ turns $6$, where the two clients can follow the network paths (i) client 1: $1\to 3 \to 6\,(\text{processing}) \to 8 \to 9$, (ii) client 2: $3\to 6 \to 5\,(\text{processing}) \to 7 \to 10 \to 11$ to benefit from cheap computation resources at node $5$ and $6$.




\section{Conclusion} \label{sec:conclusion}

In this paper, we investigated the problem of optimal cloud network control with strict deadline constraints.
We established a new queuing system to keep track of the data packets' lifetimes, on which basis we formalized the problem $\mathscr{P}_1$.
An equivalent problem $\mathscr{P}_2$ was derived, for which we provided a solution leveraging Lyapunov optimization theory.
We then took advantage of their close relationship (identical capacity region, flow space, and optimal cost value), to develop a provably near-optimal, fully distributed algorithm for $\mathscr{P}_1$, from the empirical decisions made for $\mathscr{P}_2$.
Numerical results validated the theoretical analysis and the performance gain of the proposed design over the state-of-the-art algorithm.

\ifCLASSOPTIONcaptionsoff
  \newpage
\fi

\end{document}